\documentclass[sigconf]{acmart}

\AtBeginDocument{%
  }

\setcopyright{acmlicensed}
\copyrightyear{2018}
\acmYear{2018}
\acmDOI{XXXXXXX.XXXXXXX}

\copyrightyear{2024}
\acmYear{2024}
\setcopyright{acmlicensed}\acmConference[CIKM '24]{Proceedings of the 33rd ACM International Conference on Information and Knowledge Management}{October 21--25, 2024}{Boise, ID, USA} \acmBooktitle{Proceedings of the 33rd ACM International Conference on Information and Knowledge Management (CIKM '24), October 21--25, 2024, Boise, ID, USA}
\acmDOI{10.1145/3627673.3679610} 
\acmISBN{979-8-4007-0436-9/24/10}

\usepackage{lipsum}
\usepackage{algorithm}
\usepackage{algpseudocode}
\usepackage{amsmath}
\usepackage{setspace}
\usepackage{graphicx} 
\usepackage{float} 
\usepackage{subfigure} 
\usepackage{multirow}
\usepackage{diagbox}
\usepackage{ulem}
\usepackage{color}
\usepackage{enumitem}
\usepackage[toc,page]{appendix} 
\usepackage{verbatim}
\usepackage{cleveref}
\usepackage{booktabs}
\usepackage{makecell}
\Crefformat{equation}{Equation~#2#1#3}
\usepackage{booktabs}
\usepackage{makecell}
\usepackage{microtype}

\begin{document}

\title{A Universal Sets-level Optimization Framework for Next Set Recommendation}

\author{Yuli Liu}
\affiliation{%
  \institution{Qinghai University \\
Australian National University}
  \city{Xining 810016}
  \country{China}}
\email{liuyuli012@gmail.com}

\author{Min Liu}
\affiliation{%
  \institution{Qinghai University \\
 Intelligent Computing and Application \\ Laboratory of Qinghai Province}
  \city{Xining 810016}
  \country{China}}
\email{liumin5061@gmail.com}

\author{Christian Walder}
\affiliation{%
  \institution{Google DeepMind}
  \city{Montreal}
  \country{Canada}}
\email{cwalder@google.com}

\author{Lexing Xie}
\affiliation{%
  \institution{Australian National University  
}
  \city{Canberra}
  \country{Australia}}
\email{lexing.xie@anu.edu.au}

\renewcommand{\shortauthors}{Yuli Liu, Min Liu, Christian Walder, \& Lexing Xie}

\begin{abstract}
  \underline{N}ext \underline{S}et \underline{Rec}ommendation (NSRec), encompassing related tasks such as next basket recommendation and temporal sets prediction, stands as a trending research topic. 
Although numerous attempts have been made on this topic, there are certain drawbacks: (\textit{\romannumeral1}) Existing studies are still confined to utilizing objective functions commonly found in \underline{N}ext \underline{I}tem \underline{Rec}ommendation (NIRec), such as binary cross entropy and BPR, which are calculated based on individual item comparisons; (\textit{\romannumeral2}) They place emphasis on building sophisticated learning models to capture intricate dependency relationships across sequential sets, but frequently overlook pivotal dependency in their objective functions; (\textit{\romannumeral3}) Diversity factor within sequential sets is frequently overlooked.
In this research, we endeavor to unveil a universal and \underline{S}ets-level optimization framework for \underline{N}ext \underline{S}et \underline{Rec}ommendation (SNSRec), offering a holistic fusion of diversity distribution and intricate dependency relationships within temporal sets.
To realize this, the following contributions are made: (\textit{\romannumeral1}) We directly model the temporal set in a sequence as a cohesive entity, leveraging the Structured Determinantal Point Process (SDPP), wherein the probabilistic DPP distribution prioritizes collections of structures (sequential sets) instead of individual items; (\textit{\romannumeral2}) We introduce a co-occurrence representation to discern and acknowledge the importance of different sets; (\textit{\romannumeral3}) We propose a  sets-level optimization criterion, which integrates the diversity distribution and dependency relations across the entire sequence of sets, guiding the model to recommend relevant and diversified set.
Extensive experiments on real-world datasets show that our approach consistently outperforms previous methods on both relevance and diversity.
\end{abstract}


\maketitle

\vspace{-1mm}
\section{Introduction}
\underline{N}ext \underline{I}tem \underline{Rec}ommendation (NIRec), also widely recognized as sequential recommendation \cite{kang2018self, wang2020next, adnan2021accelerating}, is a predictive model inferring the next likely item of interest from a user's previous temporal item sequence. Extending this concept, \underline{N}ext \underline{S}et \underline{Rec}ommendation (NSRec) aims to predict a subsequent set of items or products by analyzing previous set sequence. Such mechanisms are applicable in various real-world scenarios, including predicting a user's next purchase based on his/her historical transaction records or recommending a new playlist derived from the analysis of his/her daily music listening patterns over several days. Within this field of study, relevant topics encompass \underline{N}ext \underline{B}asket \underline{Rec}ommendation (NBRec) \cite{yu2016dynamic, rendle2010factorizing} and \underline{T}emporal \underline{S}ets \underline{P}rediction (TSP) \cite{hu2019sets2sets, yu2020predicting}. NBRec is derived from the pattern of consumers repeatedly buying assortments of items across different times. In contrast, temporal sets prediction is a newer and broader task that does not solely focus on item baskets but also extends to predicting set sequences in various domains, such as academic courses over semesters or daily online behaviors (clicking or adding products to shopping carts) \cite{sun2020dual}.
This approach is fundamentally a sequential sets to sequential sets learning problem, which initially does not emphasize the personalization aspect inherent in recommendation tasks \cite{chu2009personalized}. However, current research in the field \cite{yu2023continuous, yu2023predicting} frequently incorporates user preferences into the equation, thereby aligning certain temporal sets prediction tasks with the scope of NSRec. 

\begin{figure}
  \centering
  \includegraphics[width=0.89\linewidth]{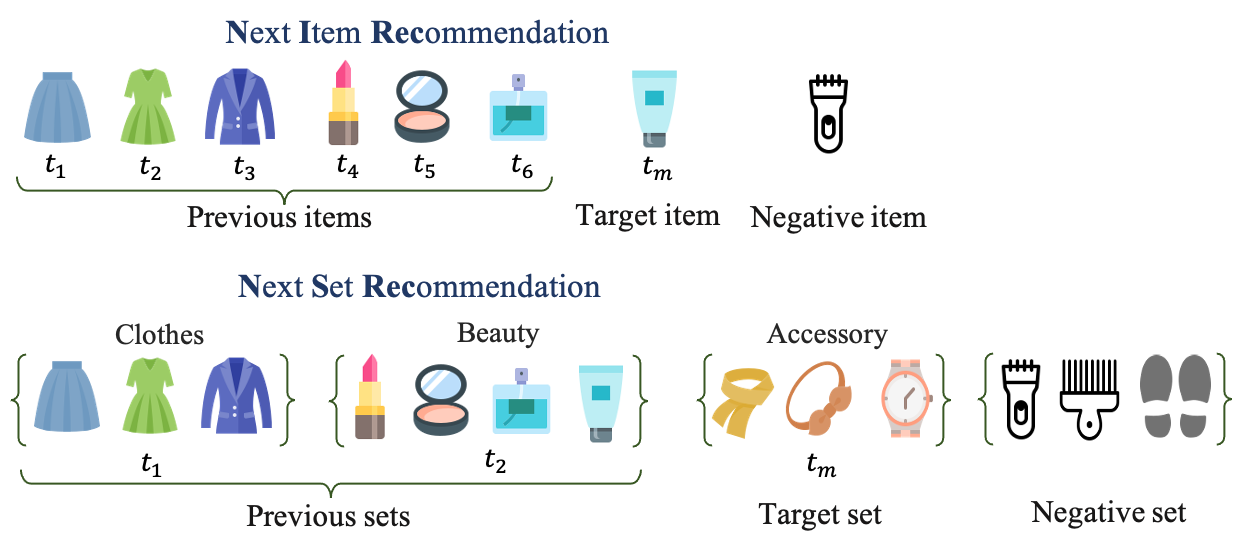}
  \vspace{-4.4mm}
  \caption{Illustration of next item/set recommendation.}
  \label{fig-SNSRec:illustration}
  \vspace{-4.5mm}
\end{figure}

\Cref{fig-SNSRec:illustration} clearly illustrates the distinction between NIRec and NSRec, with NSRec being the more complex task as it involves predicting the next set based on set-level temporal sequence, compared to NIRec’s prediction of the next item from item-level sequence. The complexity of NSRec is further amplified by the presence of an arbitrary number of items within a temporal set, necessitating sophisticated models that can discern both dependencies within sets (DinS) and across sets (DacSs). 
These advanced models typically enhance sequence processing architectures like Recurrent Neural Networks (RNNs) \cite{hu2019sets2sets, choi2016doctor, ariannezhad2022recanet} and Transformer \cite{sun2020dual, yu2022element, chou2023incorporating} to obtain item- and set-level representations.
However, it is puzzling that despite the sophisticated model designs, the development of specialized objective functions for NSRec has been overlooked. Instead, there is a tendency to employ objective functions common in NIRec, primarily pointwise loss \cite{chen2023not} (\textit{e.g.},  weighted mean square loss \cite{hu2019sets2sets}, binary cross-entropy loss \cite{sun2020dual, yu2022element, liu2020multi}, and the multi-label soft margin loss \cite{jung2021global, yu2020predicting, gurukar2022multibisage}) and pairwise  Bayesian Personalized Ranking (BPR) \cite{yu2016dynamic, romanov2023time}, which are all based on individual item comparisons. We argue that these functions may not adequately address the complexities inherent in NSRec. 
Specifically, individual item-level loss functions  presume the estimated items to be mutually independent \cite{liu2022determinantal}. For instance, the calculation of pointwise loss is achieved by contrasting the predicted values of items (in target and negative sets) against their true labels, and BPR accumulates the margins between each positive item and its negative counterpart.
As \Cref{fig-SNSRec:illustration} shows, such an approach to calculating loss neglects both the DinS  within the $Accessory$ set and the DacSs across the sequence of $Clothes \to Beauty \to Accessory$ temporal sets.

In this work, we are committed to the development of an universal optimization framework at the temporal \underline{sets} level specifically for next set recommendation. Our use of the term ‘sets' stems from the intention to capture and represent the dependencies that exist not only within the individual set but also among sets over time. An intuitive approach is to treat the temporal set as a singular element, thus enabling a shift away from individual item comparisons towards direct set entity comparisons. Representing a set, irrespective of the number of items it contains, as a whole inherently accounts for DinS and facilitates the high-level formulation of DacSs. However, three primary challenges arise in the pursuit of this vision: 
\vspace{-2.1mm}
\begin{itemize} [leftmargin=*]
\vspace{-2.1mm}
\item Considering a set as an element renders the task of discerning sequence preferences among distinct sets more complex. For example, when faced with set preferences such as $\{0.3, 0.4, 0.3\}$ and $\{0.5, 0.2, 0.3\}$, it becomes a challenge to evaluate which set holds greater importance in the context of preference learning.
\item Existing optimization methods in NSRec have centered on item-level relevance comparison, neglecting the diversity metric. Our optimization framework expands this focus to evaluate sets as a whole, thus confronting the dual challenge of modeling set sequences and maintaining their diversity. 
\item Synthesizing multiple crucial components, such as dependency, relevance, and diversity, into a comprehensive optimization framework presents a formidable challenge, particularly given the intricacies of addressing NSRec.
\end{itemize}
\vspace{-0.8mm}

To address these identified challenges, we make the following key contributions:
\vspace{-0.5mm}
\begin{itemize} [leftmargin=*]
\vspace{-0.4mm}
\item \textbf{C1}: Supplementing the sequence preference representation commonly employed in NIRec and NSRec, we introduce an innovative co-occurrence representation. This representation measures the cohesion among items within the temporal set, thereby discerning and validating the differential impact of a specific set.
\item \textbf{C2}: In line with the objective to conceptualize temporal sets as elements, our optimization framework directly models the structural properties of temporal sets by leveraging the Structured
Determinantal Point Process (SDPP), wherein the probabilistic DPP distribution prioritizes collections of structures. A structural similarity measurement approach is proposed for considering diversity in the formulation of optimization criterion.
\item \textbf{C3}: In response to the complex requirements of NSRec, we make the very first attempt to propose a specialized sets-level optimization criterion, which formulates and captures the DinS and DacSs dependencies. Guided by this criterion, we develop \textbf{a universal and \underline{S}ets-level optimization framework} for \underline{N}ext \underline{S}et \underline{Rec}ommendation (SNSRec), which promotes an informed learning process that synergizes preference and co-occurrence representations, and preserving diversity in the set sequences.
\item \textbf{C4}: SNSRec stands out for its generality  applicability and practicality, allowing mainstream NSRec methods to be deployed within it. This integration, in conjunction with SNSRec's three core components (\textbf{C1}-\textbf{C3}), leads to improvements in both accuracy and diversity of NSRec.
\end{itemize}

\vspace{-1.5mm}
\section{Preliminary}
\vspace{-1mm}
This section provides some preliminaries, including the formalization of NSRec and DPP. 

\vspace{-2mm}
\subsection{Problem Formalization}
\vspace{-1mm}

Next set recommendation can be formalized as follows: Let $\mathbb{U}=\left\{u_1, \cdots, u_{|U|}\right\}$ and $\mathbb{V}=$ $\left\{v_1, \cdots, v_{|V|}\right\}$ be the entirety of users and items (items or products), respectively. $S_t$ with an arbitrary number of interacted elements is adopted to denote a user interaction set, $S_t \subset \mathbb{V}$. Given a user's historical user-set interactions represented as a temporal sequence of sets $\boldsymbol{S} =\left\{S_1, S_2, \cdots, S_t\right\}$, the goal of NSRec is to predict the subsequent set according to the historical records, that is, 
$\hat{S}_{t+1}=f\left(\boldsymbol{S}\right).$

\vspace{-2mm}
\subsection{Determinantal Point Processes}
\label{ch:DPP}
\vspace{-1mm}
As elegant probabilistic models, determinantal point processes (DPPs) have been widely employed for diversification \cite{kulesza2012determinantal, liu2022determinantal, kulesza2010structured, gillenwater2012discovering}. 
A DPP $\mathcal{P}$ over a ground set $\mathcal{Y}=\{1,2, \ldots, M\}$, namely the item or product catalog, is a probability measure on $2^{\mathcal{Y}}$ (the set of all subsets of $\mathcal{Y}$ ). 
If $Y \subseteq \mathcal{Y}$ is a random subset drawn according to $\mathcal{P}$, the probability is $\mathcal{P}(Y) \propto \operatorname{det}\left(\mathbf{L}_Y\right)$, where $\mathbf{L} \in \mathbb{R}^{M \times M}$ is a real, positive semi-definite kernel matrix indexed by the elements of $Y$.
The marginal probability of including one element $Y_i$ is $\mathcal{P}\left(Y_i\right)=\mathbf{L}_{i i}$. 
That is, the diagonal of $\mathbf{L}$ gives the marginal probabilities of inclusion for individual elements of $\mathcal{Y}$. The probability of selecting two items $Y_i$ and $Y_j$ is $\mathrm{L}_{i i} \mathrm{L}_{j j}-\mathrm{L}_{i j}^2=\mathcal{P}\left(Y_i\right) \mathcal{P}\left(Y_j\right)-\mathrm{L}_{i j}^2$. 
The entries of the kernel $\mathbf{L}$ usually measure similarity between pairs of elements in $\mathcal{Y}$. Thus, highly similar elements are unlikely to appear together. In this way, the repulsive characteristic (\textit{i.e.}, diversity) with respect to a similarity measure is captured. 

\vspace{-1mm}
\section{Methodology}
\vspace{-1mm}
In this section, we introduce SNSRec by first modeling the temporal set as a structure, and then introduce preference and co-occurrence representations, which are used to measure the structure's quality. Finally, we present our sets-level optimization criterion. 

\vspace{-1mm}
\subsection{Structural Temporal Set Modeling}
\vspace{-0.5mm}
Reviewing the formalization of NSRec, \textit{i.e.}, forecasting the subsequent sets given previous set sequence, the basic component of this prediction task is the set instead of individual items or products, which is also the point that makes this problem apart from NIRec. This implies that standard DPPs are inadequate for the complex relations in set sequence, leading us to consider the structured determinantal point process (SDPP) \cite{kulesza2010structured} for its proficiency in modeling sets with inherent structures, such as temporal sets in this work. In SDPP case, elements of a ground set $\mathcal{Y}$ are structures, which means that we will no longer think of $\mathcal{Y}=\{1,2, \ldots, M\}$; instead, each element $Y_i \in \mathcal{Y}$ is a structure given by a collection of $R$ parts $\left(Y_i^{(1)}, Y_i^{(2)}, \ldots, Y_i^{(R)}\right)$, and each part takes a value from a finite set of $M^R$ possibilities. In the context of NSRec, each structure element is a user interaction set $S_t$ at time $t$ containing $R$ items or products in $\mathbb{V}$. Thus, a temporal set can be explicitly represented as an integrated structure. A random subset drawn from a SDPP $\mathcal{P}$ can be a temporal sequence of sets, \textit{i.e.}, $\boldsymbol{S}$. 
In this setting, we can regard NSRec as a problem of drawing salient (accurate and diversified) sets of temporal structures according to the SDPP distribution. 
SDPP directly treats the temporal set as a united element (DinS) and possesses powerful ability of capturing dependencies across sets (DacSs) incorporating notions of quality and diversity by means of global probabilistic measure, and thus naturally fits the NSRec problem. 
As a result, introducing SDPP into NSRec will push the corresponding research into a new frontier. 

However, we need to note that when the number of elements in $\mathcal{Y}$ is exponential, structured DPPs actually model a distribution over the doubly exponential number of subsets of an exponential $\mathcal{Y}$ \cite{gillenwater2012discovering}. 
An immediate challenge is that it is difficult to explicitly write down the SDPP kernel \textbf{L}.
In this work, we therefore conduct SDPP distribution calculation based on the set sequence specified SDPP kernel $\mathbf{L}^{(\boldsymbol{S})}$, \textit{i.e.}, a kernel over \textit{sequence ground set} that only involves structures (sets) of a set sequence $\boldsymbol{S}$ instead of over all structures. 


\vspace{-1.5mm}
\subsubsection{Quality Modeling}
\label{ch:Structure-Formulation}
We use the quality/diversity decomposition to define the entries of SDPP kernel $\mathbf{L}$,
\begin{equation}
L_{a b}=q(S_a) \phi(S_a)^{\top} \phi(S_b) q(S_b),
\label{euq-sdpp:Lab-1}
\end{equation}
where $q\left(S_a\right)$ is a nonnegative measure of the quality of a structure (set) $S_a$, and $\phi\left(S_a\right)$ represents a vector of diversity features so that $\phi\left(S_a\right)^{\top} \phi\left(S_b\right)$ is a measure of the similarity between structures $S_a$ and $\boldsymbol{S}_b$. 
SDPPs assume a factorization of the quality score $q\left(S_a\right)$ and similarity score $\phi\left(S_a\right)^{\top} \phi\left(S_b\right)$ into parts for allowing efficient normalization and sampling. Formally, SDPP decomposes quality multiplicatively and similarity additively:
\vspace{-1mm}
\begin{equation}
q\left(S_a\right)=\prod_{r=1}^R q\left(S_a^{(r)}\right) \quad \phi\left(S_a\right)=\sum_{r=1}^R \phi\left(S_a^{(r)}\right).
\label{euq-sdpp:weight-quality-sim}
\vspace{-1mm}
\end{equation}
For temporal sets, the part $S_a^{(r)}$ denotes the $r$th item in set $S_a$ that contains $R$ items. 

To measure the importance or salience of items and the relative strength of relationships between them, we calculate the weight of a temporal set (\textit{i.e.}, quality) for constructing SDPP structure as
\vspace{-1mm}
\begin{equation}
w(S_a)=\sum_{r=1}^R w\left(S_a^{(r)}\right)+\sum_{S_a^{(\alpha)} \in S_a, S_a^{(\beta)} \in S_a} w\left(S_a^{(\alpha)}, S_a^{(\beta)}\right),
\label{euq-sdpp:quality-weight}
\vspace{-1mm}
\end{equation}
where $\alpha \neq \beta$.
The first term denotes the accumulation of \textbf{items' importance}. For NSRec tasks, the importance of part $S_a^{(r)}$, \textit{i.e.} $w\left(S_a^{(r)}\right)$ of an item, can be represented by its predicted relevance score \textit{w.r.t.} the sequence $\boldsymbol{S}$, which is: 
\begin{equation}
\hat{\boldsymbol{y}}_i^P={\mathbf{p}^{(\boldsymbol{S})}}^{\top}{\mathbf{e}^P_i},
\label{euq-sdpp:preference-score}
\end{equation}
where $\mathbf{p}^{(\boldsymbol{S})}$ is the final sequence preference representation (\underline{detailed in} \underline{\Cref{sec-pre-r}}) of a sequence $\boldsymbol{S}$, and $\mathbf{e}^P_i$ denotes the preference embedding of item $v_i$ in $\mathbb{V}$ (\textit{i.e.}, the $r$th part $S_a^{(r)}$ in $S_a$).
The second term in \Cref{euq-sdpp:quality-weight} is used to represent the \textbf{strength of edges} in $S_a$. SDPP is originally designed to build structures of threads or paths  \cite{gillenwater2012discovering}, which means that there is an ordering in the structure. Therefore, only the strength of edges of adjacent elements in the path (structure) is considered. 
As for NSRec, items in the same set usually do not have an inherent ordering. We therefore propose to capture co-occurrence patterns of items on set level. Specifically, this work considers the cohesion between any two items of a set.
In this sense, the edge weight of any two items in a set, \textit{i.e.} $w\left(S_a^{(\alpha)}, S_a^{(\beta)}\right)$, used to represent the strength of cohesion between them,
\vspace{-0.6mm}
\begin{equation}
\hat{\boldsymbol{y}}_{\alpha \beta}^C={\mathbf{c}_{\alpha}^{(\boldsymbol{S})}}^{\top}\mathbf{c}_{\beta}^{(\boldsymbol{S})},
\label{euq-sdpp:co-score}
\vspace{-1mm}
\end{equation}
where $\mathbf{c}_{\alpha}^{(\boldsymbol{S})}$ and $\mathbf{c}_{\beta}^{(\boldsymbol{S})}$ are the final co-occurrence representations (\underline{detailed in \Cref{sec-SDPP:Co-occurrence-Representation}}) of items with respect to temporal sets sequence $\boldsymbol{S}$, learned based on co-occurrence embeddings. 



\vspace{-1.5mm}
\subsubsection{Diversity Modeling} 
We have detailed how to calculate the quality of a SDPP structure above. Here we present the approach of structural set similarity measurement ($\phi\left(S_a\right)^{\top} \phi\left(S_b\right)$), which contributes to the diversity component of NSRec. 
Aligning with \Cref{euq-sdpp:Lab-1} and \Cref{euq-sdpp:weight-quality-sim}, the set similarity function can be factorized into the item similarity function as follows  (\textbf{C2}):
\vspace{-0.5mm}
\begin{equation}
\begin{aligned}
Sim(S_a, S_b) &=\phi(S_a)^{\top} \phi(S_b) \\
&=\left(\sum_{S_a^{(\alpha)} \in S_a} \phi_\alpha\left(S_a^{(\alpha)}\right)^{\top}\right)\left(\sum_{S_b^{\left( \beta \right)} \in S_b} \phi_\beta \left(S_b^{(\beta)} \right)\right) \\
&=\sum_{\alpha, \beta} \phi_\alpha\left(S_a^{(\alpha)}\right)^{\top} \phi_\beta \left(S_b^{(\beta)} \right).
\label{euq-sdpp:diversity-score}
\end{aligned}
\vspace{-1mm}
\end{equation}
Here, $\phi_\alpha\left(S_a^{(\alpha)}\right)$ and $\phi_\alpha\left(S_a^{(\beta)}\right)$ represent $k$-dimensional features of two items belong to $S_a$ and $S_b$, respectively. This factorization indicates that the diversity measurement of two structures for each SDPP kernel entry is built by accumulating the similarity of any two items from different structures. 

To connect the common diversity concept (\textit{i.e.}, scope of categories) in recommendation fields \cite{wu2019pd, liang2021recommending, warlop2019tensorized, gartrell2016bayesian} and diversity measurement in SDPP, we propose a category-aware diverse kernel $\mathbf{K}$.
The number of items $|\mathbb{V}|$ in NSRec dataset is usually large, which means that learning a nonparametric full-rank diverse kernel $\mathbf{K}$ is computationally expensive.  
Referring to \cite{gartrell2017low}, we use the low-rank factorization of the $N \times N$ $\mathbf{K}$ matrix as: $\mathbf{K}=\mathbf{A} \mathbf{A}^{\top}$.
In this setting, a learning optimization criterion is built by maximizing the log-likelihood of sampling observed diverse subsets $T^{(+)}$ and minimizing the log-likelihood of sampling negative subsets $T^{(+)}$, 
\vspace{-0.5mm}
\begin{equation}
\ell=\sum_{\left(T^{(+)}, T^{(-)}\right) \in \mathcal{T}} \log \operatorname{det}\left(\mathbf{K}_{T^{(+)}}\right)-\log \operatorname{det}\left(\mathbf{K}_{T^{(-)}}\right), 
\label{euq-sdpp:K-objective}
\vspace{-0.5mm}
\end{equation}
where $\mathcal{T}$ denotes the collection of paired sets used for training. 
To obtain the observed diverse subsets, we select as many items belonging to different categories as possible from an observed temporal set. For example, there are four items from three categories ($c_1, c_2, c_3$) in a set $\{v_5^{c_1}, v_6^{c_2}, v_7^{c_3}, v_8^{c_3}\}$, and two diverse subsets $\{v_5^{c_1}, v_6^{c_2}, v_7^{c_3}\}$ and $\{v_5^{c_1}, v_6^{c_2}, v_8^{c_3}\}$ can be selected. For each selected diverse subset, a same-sized negative subset with randomly selected items will be provided. 
Considering the dependency among items and computational efficiency, observed temporal sets with item size in the range of 5 to 20 are used for selecting diversified subsets, and a subset should contain at least 3 items. 
Based on this, a category-aware diverse kernel $\mathbf{K}$ can be learned, which tends to make the corresponding DPP-distributed samples containing diverse categories. Namely, the items drawn based on pre-learned $\mathbf{K}$ are diversified. Reviewing the description of DPPs in \Cref{ch:DPP}, the diversity characteristic is contributed by the entries of DPP kernel that measure the similarity between pairs of elements. Since $\mathbf{K}$ facilitates sampling diversified subsets, we can directly take the corresponding entries of $\mathbf{K}$ into \Cref{euq-sdpp:diversity-score} as the similarity measurement of two items, which means that the corresponding vectors of items in $\mathbf{A}$ (low-rank factorization) represent their diversity features. 

\vspace{-1.7mm}
\subsection{Representation Learning}
\label{sec-SDPP:Preference-Representation}
\vspace{-0.6mm}


As we aim to introduce a universal optimization framework NSRec that optimizes both the novel co-occurrence representations and the common preference representation, we simply introduce the set sequence preference representation and detail exposition of our co-occurrence learning method.

\vspace{-1.5mm}
\subsubsection{Preference Representation}
\label{sec-pre-r}
\vspace{-0.5mm}
Learning \textbf{p}re\textbf{f}erence \textbf{r}epresentation (PFR) based on previous set sequence is a critical step to capture relevance between candidate items (that likely appear in next set) and users' dynamic interests. 
To learn expressive PFR for NSRec, we model preference learning on both item-level and set-level representations.
The item-level PFR $\mathbf{h}^{P}_v$ is learned based on the popular standard multi-head self-attention. Learning set-level PFR means computing a vector representation for each set in the temporal set sequence based on the preference embeddings of items in those sets. The set-level PFR $\mathbf{h}^{P}_{\boldsymbol{S}}$ is computed using multi-head attention referring to \cite{sun2020dual}. To obtain the final sequence preference representation  $\mathbf{p}^{(\boldsymbol{S})} \in \mathbb{R}^d$ \textit{w.r.t.} the temporal sets sequence $\boldsymbol{S}$ for anticipating relevance (\textbf{item importance}), we apply a gated fusion module \cite{pan2020intent, lv2019sdm, sun2020dual} to integrate preference from item level and set level. 

\vspace{-1.1mm}
\subsubsection{Co-occurrence Representation}
\label{sec-SDPP:Co-occurrence-Representation}
\vspace{-1mm}
Temporal set sequence stands apart from other sequence data, as the relationships among items within the set-level element complicate the task of anticipating next set. 
Previous studies in NSRec focus only on learning PFR of a temporal sequence \cite{hu2019sets2sets, yu2020predicting, rendle2010factorizing}, similar to above $\mathbf{p}^{(\boldsymbol{S})}$, which is also the common technique of NIRec \cite{liu2022determinantal, kang2018self}.
Two consequent \textbf{limitations} of this approach are evident: (\textit{\romannumeral1}) When we model a set as a structured element, the relevance of individual items within the set (determined by PFR) is insufficient to distinguish and acknowledge the unique importance of different sets; (\textit{\romannumeral2}) A crucial factor of the set, \textit{i.e.}, the level of cohesion among the items in it, is neglected, which is a characteristic that is specific to NSRec. 
For example, in a set where \textit{smartphone} is known with high relevance, and there are other two candidates, \textit{iPad} and \textit{screen protector}, present similar levels of relevance. This similarity makes it challenging to determine which item should be paired with the smartphone to form a set. However, by analyzing existing temporal sets, it could be found that the \textit{screen protector} more frequently co-occurs with the \textit{smartphone} in various baskets or daily activities. This indicates a stronger cohesion within \{\textit{smartphone}, \textit{screen protector}\} compared to \{\textit{smartphone}, \textit{iPad}\}. Utilizing this insight about cohesive strength enables a more effective differentiation between sets and facilitates completing or complementing an accurate set. 

To learn the cohesive strength behind set-level behaviors, we propose a new type of representation (\textbf{C1}), \textit{i.e.}, \textbf{c}o-\textbf{o}ccurrence \textbf{r}epresentation (COR), which is learned based on the co-occurrence embedding matrix $\mathbf{E}^C \in \mathbb{R}^{d \times|V|}$ for entire items. Unlike preference representation, COR is not directly associated with personalization, but provides extra evidence in the stage of computing relevance score for a candidate item via estimating the probability of the candidate co-occurring with other ones in the same set. 
In this work, we propose to learn COR using the item-oriented attention network, which adaptively learns each candidate's representation with respect to sets in the sequence \cite{zhou2018deep, sun2020dual}. 
Co-occurrence embeddings of items in a set sequence are fed into a multi-head self-attention layer to obtain the hidden item representations that consider sequence dependency, \textit{i.e.}, $\mathbf{H}^C=\left[\mathbf{h}_1^C, \cdots, \mathbf{h}_n^C\right] \in \mathbb{R}^{d \times n}$ with $n$ items in a temporal sequence. 
The co-occurrence representations for all items derived based on the sequence $\boldsymbol{S}$, from the item-oriented attention layer, can be formulated: 
\vspace{-1mm}
\begin{equation}
\mathbf{C}^{(\boldsymbol{S})}=\mathbf{W}_I^V \mathbf{H}^C \operatorname{Softmax}\left(\left(\mathbf{W}_I^K \mathbf{H}^C\right)^{\top} \mathbf{W}_I^Q \mathbf{E}^C \right),
\label{euq-sdpp:co-attention}
\vspace{-1mm}
\end{equation}
where $\mathbf{W}_I^Q, \mathbf{W}_I^K, \mathbf{W}_I^V \in \mathbb{R}^{d \times d}$ are learnable matrices;  $\mathbf{C}^{(\boldsymbol{S})} \in \mathbb{R}^{d \times|V|}$ contains sequence $\boldsymbol{S}$ specified item-level co-occurrence representations (final co-occurrence representations) for each candidate in $\mathbb{V}$; and $\mathbf{E}^C$ represents candidates' co-occurrence embeddings.  

To introduce COR learning, we can calculate the cohesion strength (\textbf{edge strength}) of any two items co-occurring in the same set based on the co-occurrence patterns of sets in training sequences (previous sets and target sets). In this setting, the learning process of COR will be guided to consider the co-occurrence patterns among candidates in entire training sets.
In the evaluation procedure, CORs of entire items with respect to a specific set sequence can be obtained from the item-oriented attention network, which will engage in ranking all candidate items for suggesting next sets. The learned CORs are supposed to offer high co-occurrence scores for item pairs that actually co-occur in the subsequent set, and thus provide extra evidence for set recommendation. 
\begin{figure*}
  \centering
  \includegraphics[width=.83\linewidth]{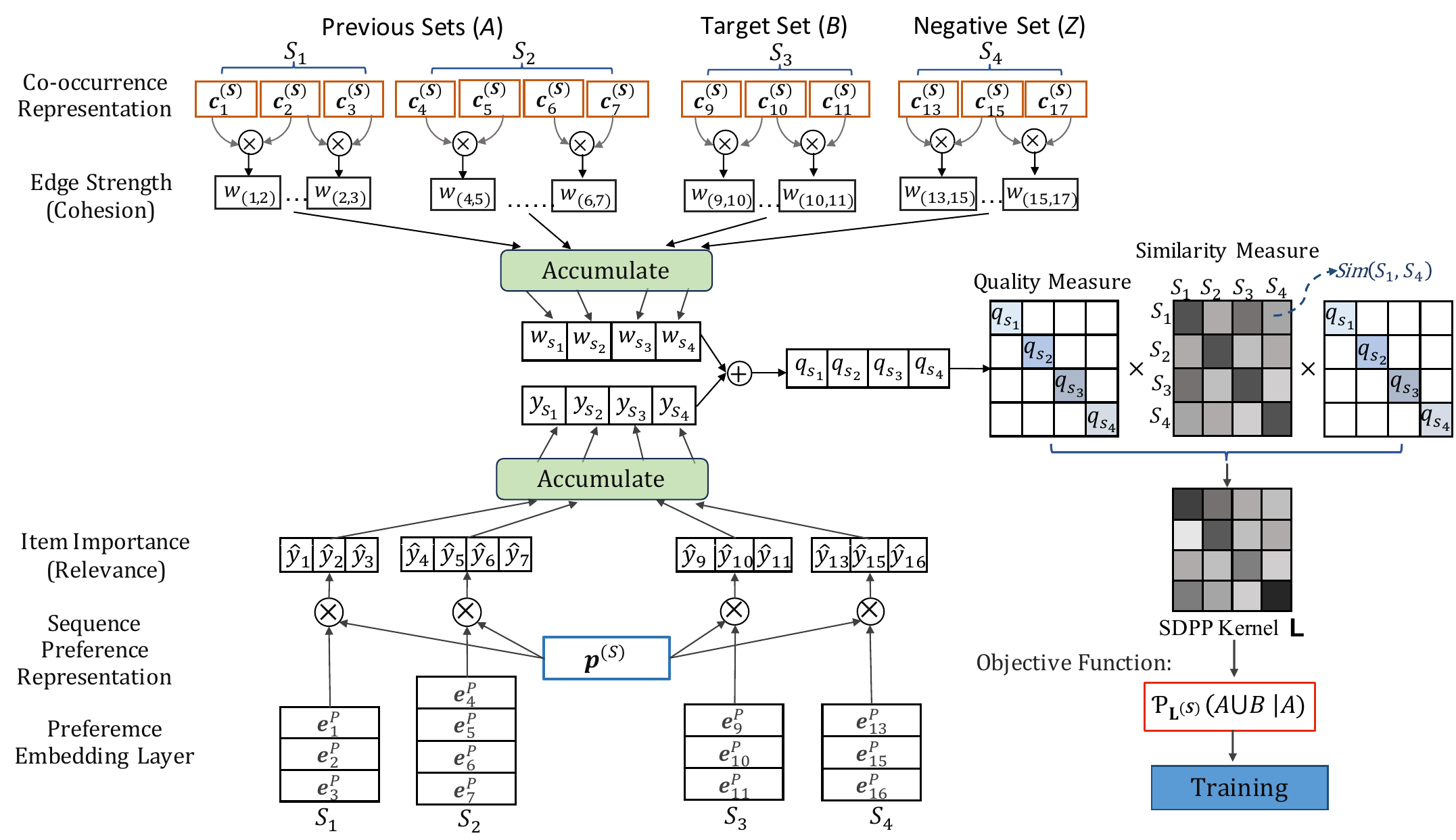}
  \vspace{-4.mm}
  \caption{An elaborate depiction for formulating the SNSRec optimization framework.}
  \label{figure-sdpp:framework}
  \vspace{-4.4mm}
\end{figure*}

\vspace{-1.7mm}
\subsection{Sets-level Optimization Criterion}
\vspace{-0.5mm}
In this work, we attempt to capture the dependency in both representation learning and optimization criterion. 
We therefore define a dedicated sets-level conditional likelihood function by defining conditional SDPP probability, 
\vspace{-1mm}
\begin{equation}
\mathcal{P}^{(\boldsymbol{S})}\left( \boldsymbol{Y}={A \cup B} \mid \boldsymbol{Y}=A\right)=\frac{\operatorname{det}\left(\mathbf{L}^{(\boldsymbol{S})}_{A \cup B}\right)}{\operatorname{det}\left(\mathbf{L}^{(\boldsymbol{S})}+\mathbf{I}_{\bar{A}}\right)}. 
\label{euq-sdpp:SDPP-likelihood}
\vspace{-0.5mm}
\end{equation}
In this equation, $\boldsymbol{Y}$ is distributed as a SDPP with temporal sets sequence $\boldsymbol{S}$ specified kernel $\mathbf{L}^{(\boldsymbol{S})}$; $A$ denotes a subset containing previous sets, and $A \cup B$ represents a subset comprised of sets of the complete training sequence (previous sets and target sets). $\mathcal{P}^{(\boldsymbol{S})}\left( \boldsymbol{Y}={A \cup B} \mid \boldsymbol{Y}=A\right)$ means the probability of if the previous sets is drawn as a SDPP subset of structures, how likely it is that sets of the complete training sequence is SDPP-distributed. 
This means that the specialized probabilistic function is naturally appropriate to the concept of NSRec, \textit{i.e.}, recommending the next sets conditioned on previous ones.
As mentioned in \Cref{ch:DPP}, we cannot afford to construct kernel $\mathbf{L}$ over all possible structures. Fortunately, the training instance in the process of NSRec learning is a definite set sequence. It means that we can form the sequence-specified ground set $\mathcal{Y}^{(\boldsymbol{S})}$ that only involves the sequence related structures, \textit{i.e.}, previous sets ($A$), target sets ($B$), and randomly selected negative sets ($Z$) that correspond to the target ones. Thus $\mathbf{L}^{(\boldsymbol{S})}$ represents the sequence-specified kernel on the ground set $\mathcal{Y}^{(\boldsymbol{S})}$. 
In the normalization constant, $\mathbf{I}_{\bar{A}}$ is the matrix with ones in the diagonal entries indexed by elements of $\mathcal{Y}^{(\boldsymbol{S})}-\bar{A}$ and zeros everywhere else.  

We then reach the optimization criterion for NSRec training by taking the logarithm of the likelihood function formulated by
\vspace{-1mm}
\begin{equation}
\mathcal{L}(\Theta)=\prod_{u \in \mathbb{U}} \prod_{{\boldsymbol{S}} \in \mathcal{D}_u} \mathcal{P}^{(\boldsymbol{S})}=\sum_{u \in \mathbb{U}} \sum_{{\boldsymbol{S}} \in \mathcal{D}_u} \log \left(\mathcal{P}^{(\boldsymbol{S})}\right),
\label{euq-sdpp:SDPP-prob}
\vspace{-1mm}
\end{equation}  
where $\boldsymbol{S}$ denotes an observed temporal sets sequence of user $u$, and $\mathcal{D}_u$ is the compilation of training sequences from $u$; $\mathcal{P}^{(\boldsymbol{S})}$ indicates the conditional likelihood (\Cref{euq-sdpp:SDPP-likelihood}) specified by a training sequence of user $u$. $\Theta$ represents model parameters including attention matrices, parameters of gated fusion module, preference embeddings, and co-occurrence embeddings. 
It can be noted that DinS and DacSs are both considered in this dedicated optimization criterion, as the corresponding calculation is not solely reliant on independent items or independent sets, but directly treats each temporal set as a SDPP structure and formulates the true set sequence as a SDPP-distributed subset (\textbf{C3}).
This criterion enables us to transform the goal of NSRec into maximizing the probability of drawing the complete set sequence as a SDPP, given the condition where previous sets are distributed as a SDPP, making the SNSRec framework rather \textit{intuitive}.
Looking at the architecture of SNSRec as depicted in \Cref{figure-sdpp:framework}, it is evident that all preparatory tasks, from learning representations to measuring diversity, are inherently aligned with this intuitive criterion. This suggests that SNSRec is purposefully designed with its goal at the core. 
\Cref{figure-sdpp:framework} also demonstrates that SNSRec is a general framework, as it can be built upon other types of representation forms for constructing SDPP structure, thus reinforcing its adaptability (\textbf{C4}).

\vspace{-1mm}
\subsubsection{Optimization}
To perform optimization for learning parameters $\Theta$ of SNSRec, we can maximize the log-likelihood 
\vspace{-1mm}
\begin{equation}
\mathcal{L}(\Theta) =\sum_{u \in \mathbb{U}} \sum_{\boldsymbol{S} \in \mathcal{D}_u} \left[ \log \left(\operatorname{det}\left(\mathbf{L}^{(\boldsymbol{S})}_{A\cup B}(\Theta)\right)\right) \\
- \log \left(\operatorname{det}\left(\mathbf{L}^{(\boldsymbol{S})}(\Theta)+ \mathbf{I}_{\bar{A}}\right)\right) \right].
\label{euq-sdpp:log-maxi}
\vspace{-1mm}
\end{equation}
Gradient-based learning methods, such as 
gradient descent and stochastic gradient descent, provide rigorous foundation for optimization because of their theoretical guarantees \cite{affandi2014learning}, but require knowledge of the gradient of $\mathcal{L}(\Theta)$. In the discrete DPP setting, this gradient can be computed straightforwardly, and we provide the computation process as follows
\vspace{-1mm}
\begin{equation}
\begin{aligned}
 \frac{\partial \mathcal{L}(\Theta)}{\partial \Theta}  &= 
 \sum_{u \in \mathbb{U}} \sum_{S \in \mathcal{D}_u} \operatorname{tr} \left(\mathbf{L}_{A \cup B}(\Theta)\right)^{-1} \frac{\partial \mathbf{L}_{A \cup B}(\Theta)}{\partial  \Theta}  \\ & - \sum_{u \in \mathbb{U}} \sum_{S \in \mathcal{D}_u} \operatorname{tr} \left(\mathbf{L}(\Theta)+\mathbf{I}_{\bar{A}}\right)^{-1} \frac{\partial  \mathbf{L}(\Theta)}{\partial  \Theta}. 
 \label{euq-sdpp:gradient-theta}
\end{aligned}
\vspace{-1mm}
\end{equation}
For brevity, we drop the superscript ($\boldsymbol{S}$) of the sequence specified SDPP kernel. In order to convert the weight function defined in \Cref{euq-sdpp:quality-weight} to the multiplicative form, a simple log-linear model is used. Consequently, the entry of SDPP kernel is  formulated as
\begin{equation}
L_{a b}= \exp \left(w(S_a)\right) Sim (S_a, S_b) \exp \left(w(S_b)\right). 
 \label{euq-sdpp:Lab-2}
\end{equation}

As a structure is regarded as an element in SDPP, we can directly compute the gradient with respect to the quality of an element $S_a$
\begin{equation}
\begin{aligned}
\frac{\partial  L_{ab}}{\partial q_a} & = \exp \left(q_a\right) Sim (S_a, S_b) \exp \left(q_b\right).
 \label{euq-sdpp:gradient-qa}
\end{aligned}
\end{equation}
$q_a$ is used to denote the quality of structure $S_a$.  

In \Cref{euq-sdpp:Lab-2}, we derive similarity measure $Sim{(S_a, S_b)}$ using diverse kernel $\mathbf{K}$, as outlined in \Cref{euq-sdpp:diversity-score}. This kernel has been pre-learned and does not require further learning during the SNSRec optimization process.  
Furthermore, the qualities of the structures ($S_a$ and $S_b$) are computed by accumulating the weights of their elements as shown in \Cref{euq-sdpp:quality-weight}, a simple process that ensures all components are differentiable and thus allows for the computation of gradients at each location \cite{affandi2014learning}. 
We can simplify the computation of the gradient of $q_a$ with respect to its parameters within the SDPP kernel by applying the chain rule to decompose the quality weight for calculating gradients of each component \cite{chao2015large} .
These parameters, which include parameterized preference and co-occurrence representations in \Cref{euq-sdpp:preference-score} and \Cref{euq-sdpp:co-score}, respectively, form the SDPP optimization criterion, denoted as $\theta_a$, and $\frac{\partial q_a}{\partial \theta_a}=\frac{\partial w\left(S_a\right)}{\partial \theta_a}$. 
Substitute decomposed gradients back into the previous gradient formula \Cref{euq-sdpp:gradient-theta}, and we can further receive the final results.
That is, the SDPP-based sets-level optimization criterion is differentiable with respect to its parameters.
To obtain the gradients of entire model parameters, the attractive backpropagation algorithm \cite{rosenblatt1961principles} can be applied. 
Finally, the gradient-based algorithms are allowed to be applied for optimization. In experiments, Adam (an variant of stochastic gradient descent) \cite{kingma2014adam} is used. 

\vspace{-1mm}
\subsubsection{Prediction}
Now we can introduce how to synthetically apply the learned co-occurrence representations $\mathbf{C}^{(\boldsymbol{S})}$ of entire items and the sequence preference representation $\mathbf{p}^{(\boldsymbol{S})}$ for predicting the subsequent set of temporal sets sequence $\boldsymbol{S}$. For an item $v_i$, the final predicted score for evaluation is 
\vspace{-1mm}
\begin{equation}
\begin{aligned} \hat{\boldsymbol{y}}_i = (1-\lambda) {\mathbf{p}^{(\boldsymbol{S})}}^{\top} \mathbf{e}_i^P + \lambda \frac{1}{|V|}\sum\limits_{n=1}^{|V|} {\mathbf{c}_i^{(\boldsymbol{S})}}^{\top} \mathbf{c}_n^{(\boldsymbol{S})}.
 \label{euq-sdpp:evaluation}
\end{aligned}
\vspace{-1mm}
\end{equation}
The co-occurrence score of $v_i$ in the subsequent set given previous sequence is measured by averaging the co-occurrence strength between $v_i$ and any one of other items in $\mathbb{V}$. In this setting, we expect the cohesion score for a target item $v_t$ in the subsequent set to be relatively high. This is because CORs are learned to reflect co-occurrence patterns, thereby enhancing the cohesion strength between $v_t$ and other target items of the next set. These cohesion scores are aggregated as shown in \Cref{euq-sdpp:evaluation}. 
Furthermore, we use the parameter $\lambda$ to balance the contribution of co-occurrence and preference scores when evaluating predictions.

\vspace{-0.5mm}
\section{Experiments}
\vspace{-0.5mm}
This section comprehensively evaluates and compares our universal optimization framework, SNSRec. 

\begin{table}[tp]
\vspace{-1mm}
\centering
  \fontsize{8}{9}\selectfont
  \caption{Statistics of the datasets.}
  \vspace{-3.7mm}
  \setlength{\tabcolsep}{3mm}{
    \begin{tabular}{ccccccc}
    \hline
    Dataset &\#Users&\#Items&\#Sets&\#Categories \\
    \textit{TaoBao}& 12.0K & 14.2K & 85.9K  & 68 \\ 
    \textit{JD-buy}& 13.5K & 19.6K & 76.8K & 77 \\ 
    \textit{JD-add}& 3.4K & 12.1K & 16.1K & 75\\ 
    \textit{TaFeng}& 10.3K  &8.7K &73.3K & 650\\ 
    \hline
    \end{tabular}}
    \vspace{-3mm}
    \label{table-sdpp:datasets}
\end{table}

\vspace{-1mm}
\subsection{Experimental Settings}
\vspace{-0.6mm}
 \subsubsection{Datasets} 
Experiments are conducted on four real-world datasets from three application scenarios. These datasets are commonly found in two topics (NBRec and TSP) related to NSRec, 
(\textit{i}) \textbf{TaoBao}\footnote{https://tianchi.aliyun.com/dataset/dataDetail?dataId=649} denotes a public online e-commerce dataset provided by Ant Financial services. It contains the interactions about which items are purchased or clicked by each user with the timestamp. The items purchased in the same day are treated as temporal sets \cite{sun2020dual, yu2023predicting} or baskets \cite{romanov2023time, che2019inter};
(\textit{ii}) \textbf{JingDong}\footnote{https://jdata.jd.com/html/detail.html?id=} is comprised of user action records, including browsing, purchasing, following, commenting and adding to shopping carts, which are widely used in TSP  \cite{yu2020predicting, yu2023continuous} and NBRec \cite{liu2020next}. Two datasets are extracted from JingDong, that is, \textbf{JD-buy} and \textbf{JD-add}, which contain purchasing and adding to shopping carts records respectively. We treat all the items purchased/added in the same day as a set;
(\textit{iii}) \textbf{TaFeng}\footnote{https://www.kaggle.com/chiranjivdas09/ta-feng-grocery-dataset} dataset contains the transactions about which items are bought by each customer in each basket with the time stamp. This dataset is commonly used in both of NBRec and TSP tasks. We treat products purchased in the same day by the same customer as a set. 
The statistics of all the datasets are shown in \Cref{table-sdpp:datasets}.

For each user in dataset, we can get a series of training instances. As indicated in \Cref{figure-sdpp:framework}, a training set instance (\textit{i.e.}, the set sequence specified ground set $\mathcal{Y}^{(\boldsymbol{S})}$) for the SDPP optimization criterion is comprised of previous sets, target sets, and negative sets, whose sizes (number of sets) are denoted using $A$, $B$, and $Z$, respectively in subsequent discussions. If a temporal set contains only one item, the edge weight is omitted. 
Following previous TSP or NBRec work \cite{yu2020predicting, sun2020dual, romanov2023time, fouad2022efficient}, each dataset is divided into three parts. The last set, the second last set, and the remaining sets of each user are used for testing, validation, and training, respectively.

\vspace{-1.5mm}
 \subsubsection{Baselines}
To validate the superiority of our framework, multiple state-of-the-art models in NBRec and TSP fields built based on different techniques are selected as baselines, which are related to this work and are competitive. \textbf{DREAM} \cite{yu2016dynamic} is a RNN-based method that uses a pooling layer to represent a basket for next basket recommendation. \textbf{CDSL} \cite{liu2022determinantal} uses the standard DPP likelihood as the loss function of sequential recommendation. To implement CDSL for NSRec, we treat the preference representation and kernel $\mathbf{K}$ of SNSRec as the quality and diversity components of DPP in CDSL. \textbf{FPMC} \cite{rendle2010factorizing} designs a hybrid model to take temporal dependencies between items and user’s general interests into account.
\textbf{DIN} \cite{zhou2018deep} proposes a DNN-based method used for click-though rate prediction, which uses attention mechanism to learn the representations of candidate items \textit{w.r.t.} historical behaviors. 
\textbf{Sets2Sets} \cite{hu2019sets2sets} is the seminal work for TSP. The average pooling operation is used to obtain the representations for sets. \textbf{DSNTSP} \cite{sun2020dual} designs a attention-based network to learn item-level and set-level representations of user sequences for TSP. \textbf{ETGNN} \cite{yu2022element} is a competitive TSP work. It uses graph neural network to propagate set-level information for obtaining collaborative signals for temporal sets. \textbf{SFCN} \cite{yu2023predicting} anticipates the next set (transactions or daily behaviors) based on specialized neural networks. 

\vspace{-1.5mm}
\subsubsection{Configuration}
We implement our optimization framework using PyTorch on a NVIDIA Quadro P2000 GPU. Adam is adopted as the optimizer in our experiments. The number of heads ($H$) in our representation learning module is set to 4,  and 4 trainable queries ($K$) are used. 
We tune the hyper parameters in all the compared methods with grid search to choose the best performance for comparison.
To keep the comparison fair, the dimension of embeddings in baselines is set to 128, and the PFR and COR are both represented by 64-dimensional vectors. 

\vspace{-1.5mm}
\subsubsection{Evaluation Metrics}
To comprehensively evaluate the quality and diversity of SNSRec, three groups of evaluation metrics are adopted: (\textit{i}) \textbf{Accuracy}. We evaluate the accuracy of set prediction using Recall@$N$ and NDCG@$N$. To check the relevance of the predicted set in different ranking positions of the ranking list, we evaluate $N=20$ and $N=50$; 
(\textit{ii}) \textbf{Diversity}. The Category Coverage (CC) \cite{wu2019pd, puthiya2016coverage} and Intra-List Distance (ILD) \cite{puthiya2016coverage, zhang2008avoiding} are prevalent and intuitive diversity evaluation metrics, which are appropriate to the diversity concept in recommendation area. 
(\textit{iii}) \textbf{Trade-off}. To evaluate the balance performance between accuracy and diversity, the harmonic metric F-score (F1) \cite{cheng2017learning, liu2020diversified} is used. F1@$N$ = $2 \times \text{Q@}N \times   \text{D@}N / ( \text{Q@}N +  \text{D@N}$), where Q@$N$ and D@$N$ are the averaged value of accuracy and diversity metrics of Top-$N$ list.

\vspace{-1.7mm}
\subsection{Experimental Results}

\vspace{-0.8mm}
\subsubsection{Overall Comparison}

\begin{table*}[tp]
\centering
  \fontsize{7.3}{7.7}\selectfont
  \caption{Overall performance comparisons. The best performing model is boldfaced, while the second best is marked with an asterisk. Improvements over baselines are statistically significant with $p<0.01$.}
  \vspace{-3.5mm}
  \setlength{\tabcolsep}{2.6mm}{
    \begin{tabular}{llllllllll|cc}
    \toprule[1pt]
Dataset&Metric&DREAM&CDSL&FPMC&\ \ DIN&Sets2Sets&ETGNN&DSNTSP&SFCN&\ SNSRec&\textit{Improv} (\%) \cr\midrule[0.8pt]
    \multirow{10}{*}[-0pt]{TaoBao} 
    &Recall@20 &0.1053 &0.1306&0.1032&0.1160&0.1286&0.1503&0.1457&0.1544$^{\boldsymbol{*}}$&\textbf{0.1716}&11.14   \\
    &Recall@50 &0.1526&0.1855&0.1639&0.1795&0.1920&0.2138&0.2046&0.2152$^{\boldsymbol{*}}$&\textbf{0.2549}&18.45 \\
    &NDCG@20 &0.0885&0.1092&0.0850&0.0915&0.1001&0.1170$^{\boldsymbol{*}}$&0.1095&0.1084&\textbf{0.1247}&6.58 \\
    &NDCG@50 &0.1033&0.1184&0.1007&0.1090&0.1211&0.1294&0.1363$^{\boldsymbol{*}}$&0.1343&\textbf{0.1490}&9.32 \\
    \cmidrule(r){2-12}
    &CC@20 &0.0905&0.1253$^{\boldsymbol{*}}$&0.0912&0.1001&0.0986&0.1104&0.1187&0.1126&\textbf{0.1360}&8.54 \\
    &CC@50 &0.1921&0.2201$^{\boldsymbol{*}}$&0.1938&0.1993&0.1953&0.1870&0.2188&0.1924&\textbf{0.2278}&3.50 \\
    &ILD@20 &0.6613&0.6668$^{\boldsymbol{*}}$&0.6028&0.6435&0.6492&0.6190&0.6105&0.6400&\textbf{0.6957}&4.33 \\ 
    &ILD@50 &0.6870&0.7012$^{\boldsymbol{*}}$&0.6531&0.6720&0.6890&0.6557&0.6617&0.6705&\textbf{0.7299}&4.09 \\ 
    \cmidrule(r){2-12}
    &F1@20& 0.1541&0.1841&0.1481&0.1622&0.1751&0.1956$^{\boldsymbol{*}}$&0.1890&0.1948&\textbf{0.2185}&11.68 \\
    &F1@50& 0.1982&0.2285&0.2016&0.2167&0.2312&0.2439&0.2458&0.2487$^{\boldsymbol{*}}$&\textbf{0.2841}&14.21\\
    \midrule[0.8pt]
    \multirow{10}{*}[-0pt]{JD-add} 
    &Recall@20 &0.1047&0.1107&0.0973&0.1027&0.1085&0.1306$^{\boldsymbol{*}}$&0.1291&0.1250&\textbf{0.1640}&25.57 \\ 
    &Recall@50 &0.1416&0.1562&0.1332&0.1407&0.1476&0.1734$^{\boldsymbol{*}}$&0.1616&0.1705&\textbf{0.1910}&10.15 \\
    &NDCG@20 &0.0810&0.0957&0.0737&0.0794&0.0843&0.1027&0.1160$^{\boldsymbol{*}}$&0.1085&\textbf{0.1430}&23.28 \\ 
    &NDCG@50 &0.0902&0.1043&0.0850&0.0918&0.0967&0.1165&0.1252$^{\boldsymbol{*}}$&0.1190&\textbf{0.1521}&21.49 \\
    \cmidrule(r){2-12}
    &CC@20 &0.1105&0.1118$^{\boldsymbol{*}}$&0.0937&0.1067&0.0918&0.1085&0.1112&0.1090&\textbf{0.1125}&6.30 \\ 
    &CC@50 &0.1824&0.1976&0.1780&0.1967&0.1952&0.1812&0.1990$^{\boldsymbol{*}}$&0.1889&\textbf{0.2002}&0.60 \\
    &ILD@20 &0.6220&0.6384$^{\boldsymbol{*}}$&0.6152&0.6380&0.6200&0.6214&0.6352&0.6209&\textbf{0.6535}&2.37 \\ 
    &ILD@50 &0.6434&0.6502&0.6437&0.6513$^{\boldsymbol{*}}$&0.6503&0.6499&0.6510&0.6431&\textbf{0.6918}&6.22 \\ 
    \cmidrule(r){2-12}
    &F1@20 &0.1481&0.1619&0.1378&0.1463&0.1517&0.1768&0.1845$^{\boldsymbol{*}}$&0.1769&\textbf{0.2192}&18.78 \\
    &F1@50 &0.1810&0.1993&0.1724&0.1825&0.1895&0.2149$^{\boldsymbol{*}}$&0.2144&0.2147&\textbf{0.2478}&15.29 \\
    \midrule[0.8pt]
    \multirow{10}{*}[-0pt]{JD-buy} 
    &Recall@20 &0.3122&0.3106&0.3015&0.3298&0.3158&0.3325&0.3406$^{\boldsymbol{*}}$&0.3374&\textbf{0.3812}&11.92 \\ 
    &Recall@50 &0.3460&0.3779&0.3528&0.3829&0.3781&0.4153$^{\boldsymbol{*}}$&0.4090&0.4146&\textbf{0.4634}&11.58 \\
    &NDCG@20 &0.2157&0.2510&0.2172&0.2453&0.2312&0.2637$^{\boldsymbol{*}}$&0.2576&0.2619&\textbf{0.2804}&6.33 \\ 
    &NDCG@50 &0.2465&0.2683&0.2350&0.2627&0.2542&0.2759&0.2825$^{\boldsymbol{*}}$&0.2702&\textbf{0.3080}&9.03 \\
    \cmidrule(r){2-12}
    &CC@20 &0.0942&0.0954$^{\boldsymbol{*}}$&0.0929&0.0913&0.0950&0.0914&0.0947&0.0910&\textbf{0.0974}&2.10 \\ 
    &CC@50 &0.1802&0.1936$^{\boldsymbol{*}}$&0.1783&0.1896&0.1873&0.1826&0.1910&0.1815&\textbf{0.1996}&3.10 \\
    &ILD@20 &0.6411&0.6430&0.6381&0.6443$^{\boldsymbol{*}}$&0.6276&0.6410&0.6289&0.6340&\textbf{0.6712}&4.18 \\ 
    &ILD@50 &0.6926&0.6977$^{\boldsymbol{*}}$&0.6743&0.6831&0.6683&0.6791&0.6705&0.6838&\textbf{0.7180}&2.91 \\ 
    \cmidrule(r){2-12}
    &F1@20 &0.3073&0.3190&0.3034&0.3228&0.3113&0.3287$^{\boldsymbol{*}}$&0.3275&0.3281&\textbf{0.3555}&8.18 \\
    &F1@50 &0.3529&0.3746&0.3479&0.3711&0.3636&0.3835&0.3836$^{\boldsymbol{*}}$&0.3823&\textbf{0.4191}&9.25 \\
    \midrule[0.8pt]
    \multirow{10}{*}[-0pt]{TaFeng} 
    &Recall@20 &0.0593&0.0682&0.0617&0.0652&0.0648&0.0718$^{\boldsymbol{*}}$&0.0676&0.0703&\textbf{0.0814}&13.37 \\ 
    &Recall@50 &0.0985&0.1095&0.0979&0.1008&0.1082&0.1103&0.1156$^{\boldsymbol{*}}$&0.1123&\textbf{0.1264}&9.34 \\
    &NDCG@20 &0.0512&0.0548&0.0516&0.0497&0.0509&0.0562$^{\boldsymbol{*}}$&0.0531&0.0527&\textbf{0.0630}&12.10 \\ 
    &NDCG@50 &0.0609&0.0655&0.0617&0.0622&0.0641&0.0691$^{\boldsymbol{*}}$&0.0680&0.0667&\textbf{0.0754}&9.12 \\
    \cmidrule(r){2-12}
    &CC@20 &0.0210&0.0221$^{\boldsymbol{*}}$&0.0206&0.0220&0.0217&0.0197&0.0213&0.0218&\textbf{0.0223} &0.90 \\ 
    &CC@50 &0.0452&0.0489&0.0484&0.0497$^{\boldsymbol{*}}$&0.0475&0.0464&0.0491&0.0420&\textbf{0.0500}&0.60 \\
    &ILD@20 &0.5771&0.5938$^{\boldsymbol{*}}$&0.5692&0.5837&0.5710&0.5781&0.5812&0.5796&\textbf{0.6153}&3.62 \\ 
    &ILD@50 &0.6074&0.6126$^{\boldsymbol{*}}$&0.5934&0.6091&0.5970&0.6098&0.6122&0.6101&\textbf{0.6416}&4.73 \\ 
    \cmidrule(r){2-12}
    &F1@20 &0.0933&0.1025&0.0950&0.0966&0.0968&0.1054$^{\boldsymbol{*}}$&0.1006&0.1021&\textbf{0.1177}&11.68 \\
    &F1@50 &0.1281&0.1384&0.1278&0.1307&0.1360&0.1409&0.1437$^{\boldsymbol{*}}$&0.1404&\textbf{0.1562}&8.71 \\
    \bottomrule[1pt]
    \end{tabular}}
    \label{table-sdpp:all-performance}
    \vspace{-3.1mm}
\end{table*}

The overall comparison is reported in \Cref{table-sdpp:all-performance}. The relative improvement (\textit{Improv}) of SNSRec over the highest performance out of baselines is also listed. The main observations obtained from the overall comparison are:
\begin{itemize}[leftmargin=*]
\vspace{-0.8mm}
\item In most cases, the improvements of quality performance (Recall and NDCG) over the best baselines are greater than 10\% (some cases even achieve $>20\%$, \textit{e.g.},  NDCG@50 on JD-add). 
The convincing outcomes of our experiments affirm the value of our primary innovations, including direct modelling temporal sets as structures, the formulation of a specialized optimization criterion, and the exploration of DinS and DacSs. These results provide compelling evidence that our proposed SNSRec contributes significantly to improve performances (\textbf{C3}) in NSRec.
\item SNSRec exhibits superior diversity performance (CC and ILD) compared to the baselines (including CDSL that considers the diversity of individual items). This highlights the effectiveness of utilizing set-level diversity measurements for NSRec tasks (\textbf{C2}). Notably, promoting the diversity of the subsequent set holds significant practical value. For instance, in a real-world application scenario such as a retail store, diversifying predictions enables the store to offer bundles with a variety of products, thereby increasing consumers' potential purchasing desire.
\item Our method exhibits advantages in both quality and diversity when comparing the overall metric F1. This demonstrates the comprehensive effectiveness of our approach in balancing the relevance and diversity aspects of predictions. 
\vspace{-1mm}
\end{itemize} 


\vspace{-1.4mm}
\subsubsection{Generality Study \textnormal{(\textbf{C4})}}
\begin{figure}
  \centering
  \includegraphics[width=.93\linewidth]{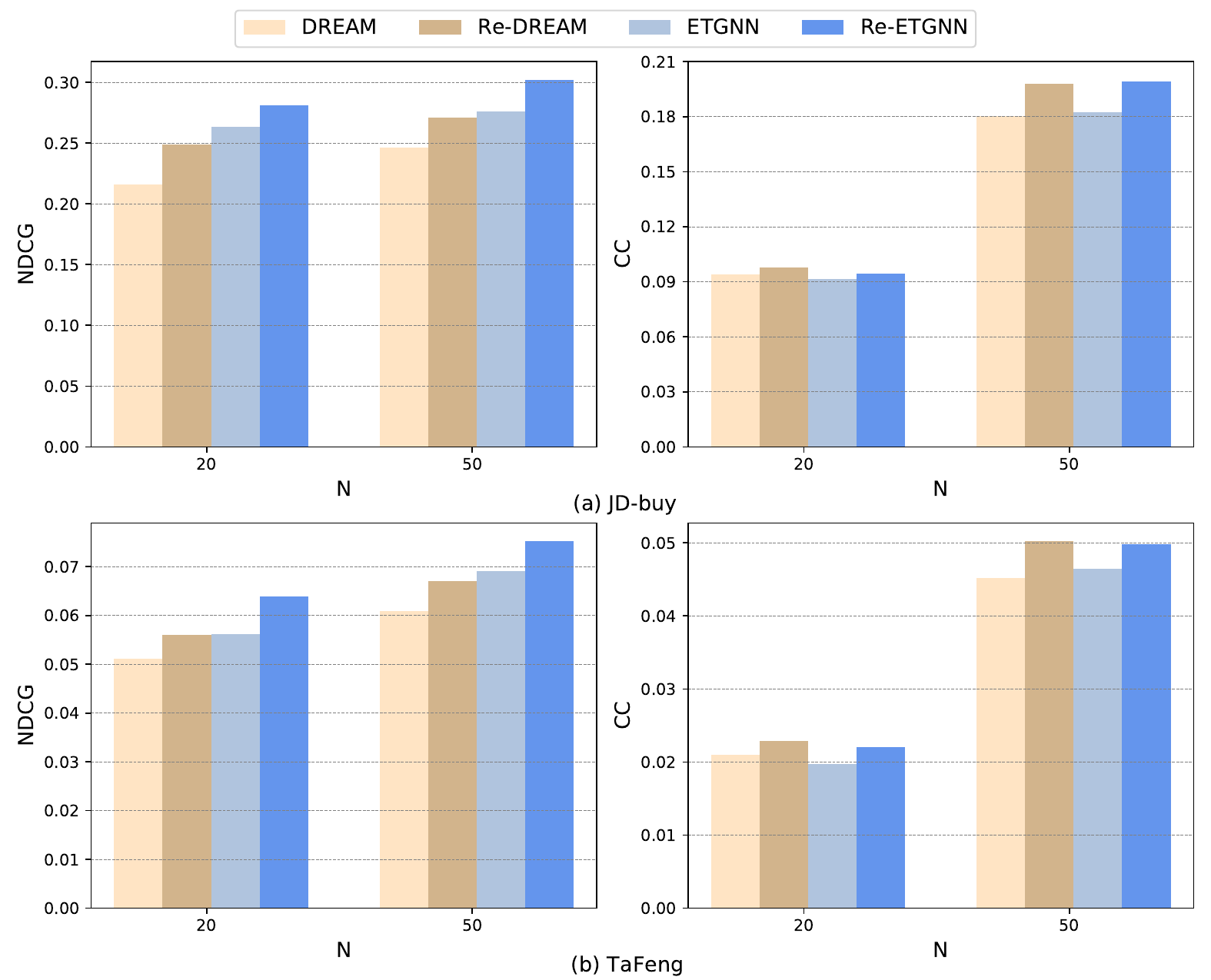}
        \vspace{-3mm}
  \caption{Performance comparison between representative  baselines and their SNSRec reworked counterparts.}
\label{figure-sdpp:reworked-models}  
    \vspace{-5mm}
\end{figure}

To validate the universality of SNSRec, we deploy original NSRec models into it. Specifically, we combine our co-occurrence learning with an existing NSRec model (\textit{i.e.}, set-level sequence preference representation learning), coupled with the pre-learned diverse kernel to represent sets as SDPP structures, and replace the original model's loss function with our specially designed sets-level optimization criterion, thereby forming a reworked model. We aim to integrate SNSRec with the most popular sequence learning algorithms (attention-based, RNN-based, and neural graph-based methods) to underscore its adaptability and wide applicability in the realm of modern techniques. Given that our preference learning primarily employs attention-based methods with notable effectiveness, we select RNN-based DREAM and graph-based ETGNN as the other two representative models. \Cref{figure-sdpp:reworked-models} illustrates the comparison between the original models and  and their versions integrated into the SNSRec framework (denoted by ‘Re-'). 
The results indicate a significant enhancement in recommendation performance for both representative models, thereby substantiating the generality and practicality of SNSRec.


\begin{figure}
\centering
    \subfigure[Top-20 on TaoBao]{
    \includegraphics[width=0.45\linewidth]{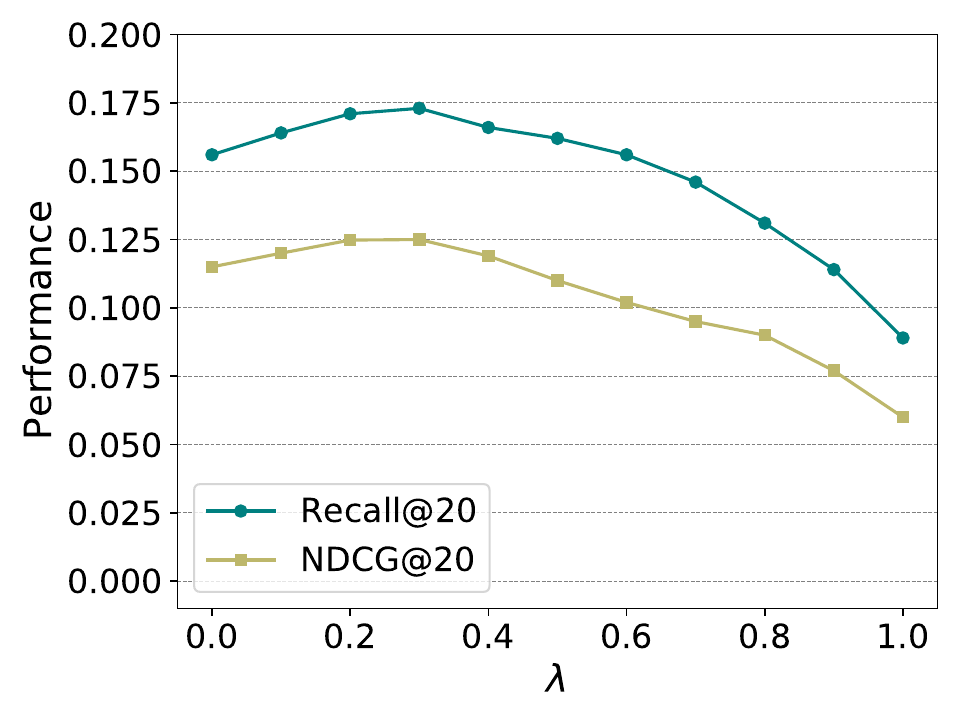}}
    \hfill
    \subfigure[Top-50 on TaoBao]{
    \includegraphics[width=0.45\linewidth]{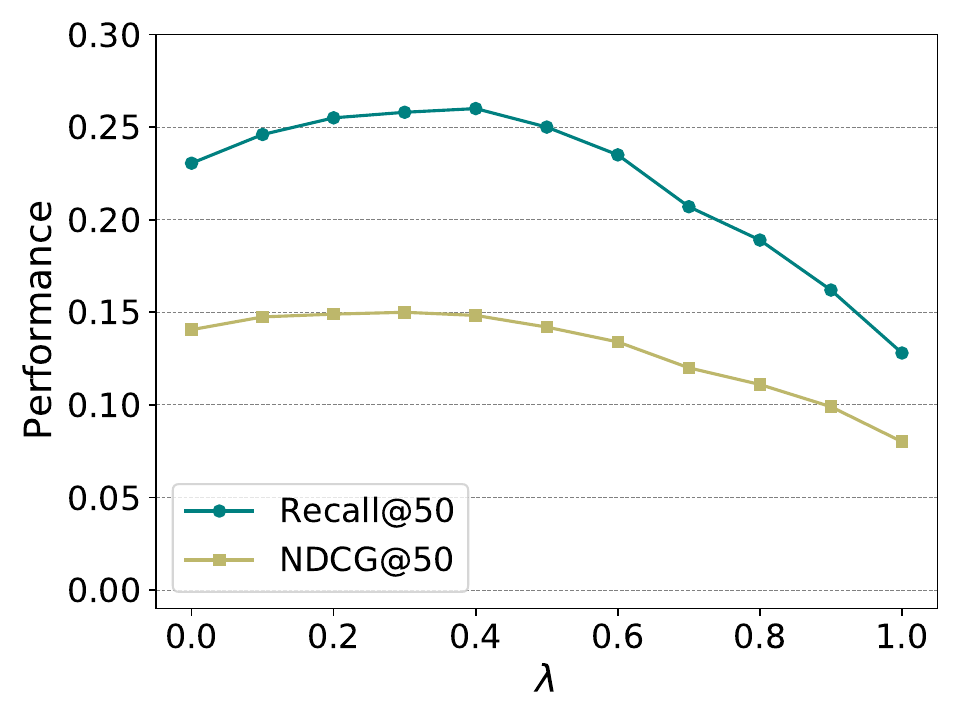}} \\
    \subfigure[Top-20 on JD-add]{
    \includegraphics[width=0.45\linewidth]{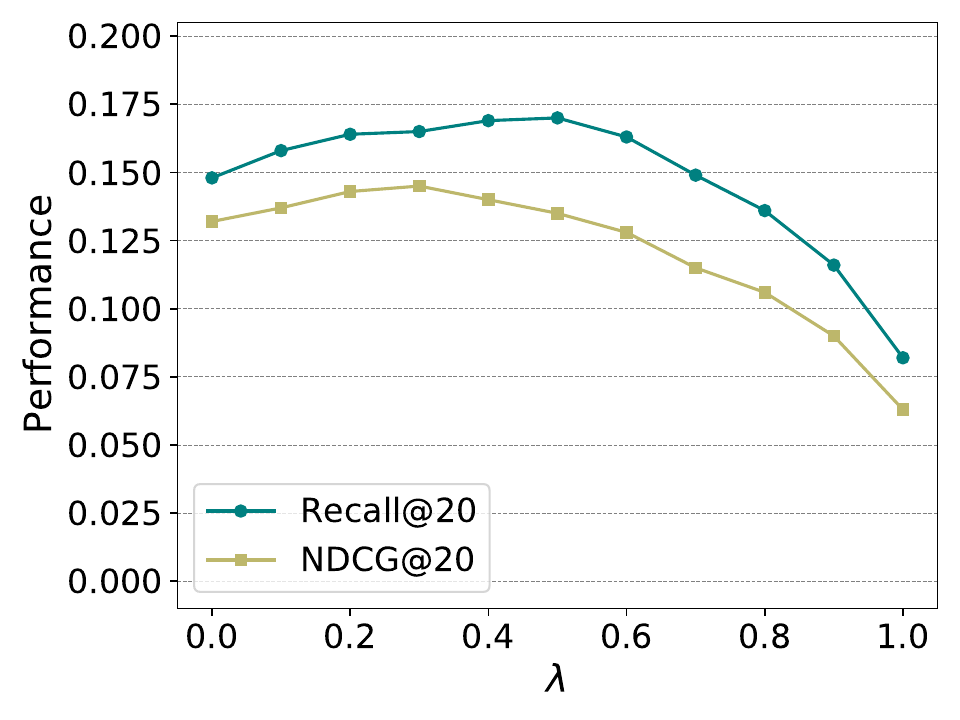}}
    \hfill
    \subfigure[Top-50 on JD-add]{
    \includegraphics[width=0.45\linewidth]{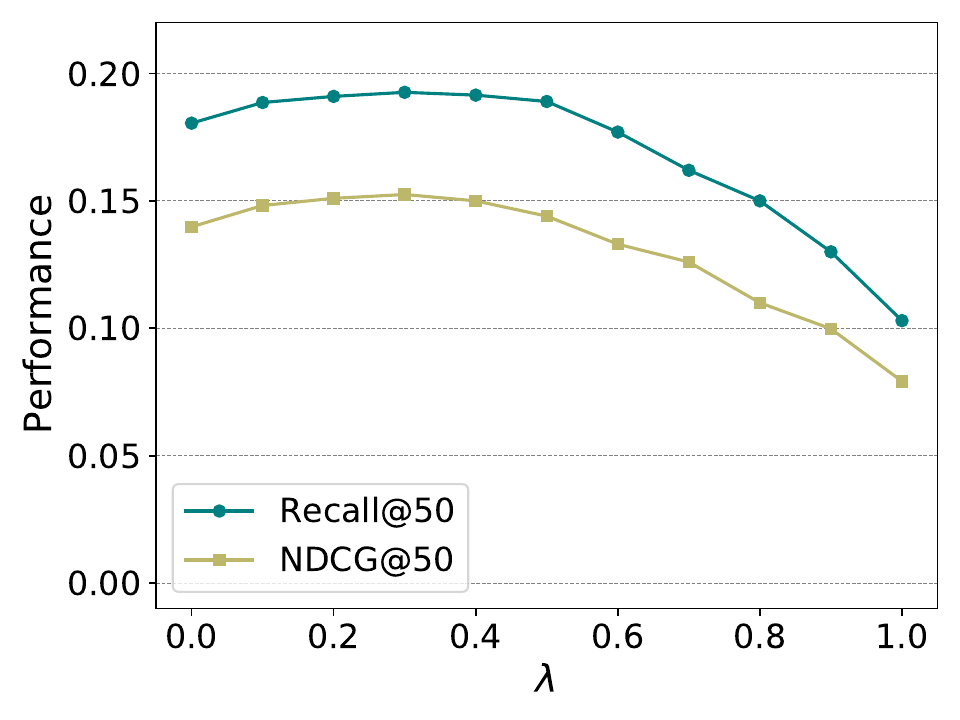}}
    \vspace{-4mm}
\caption{Performance trends with parameter $\lambda$.}
\vspace{-5mm}
\label{figure-sdpp:performance-lambda}
\end{figure}

\vspace{-1.4mm}
\subsubsection{Representations Analysis}
We conduct experiments under specialized settings to respectively analyze the importance of preference and co-occurrence representations.
\Cref{figure-sdpp:performance-lambda} shows the trends of quality performance (Recall and NDCG) with parameter $\lambda$ (introduced in \Cref{euq-sdpp:evaluation}) on TaoBao and JD-add.
We describe two extreme cases at first, \textit{i.e.}, $\lambda=0$ and $\lambda=1$.  
In the case of $\lambda=1$, only co-occurrence score is used for calculating prediction score, and thus the learning of preference representation is deprecated from SNSRec. 
When $\lambda=0$, the co-occurrence score is not activated, and the learning of co-occurrence representations is not involved in the training process, \textit{i.e.}, the weight of a set (structure) is represented only using relevance.
These two extreme settings can serve as a form of ablation study.
We aim to use \Cref{figure-sdpp:performance-lambda} to analyze the following two aspects: (i) \textbf{Importance of Co-occurrence}. When $\lambda$ equals 0 (without using co-occurrence), SNSRec does not reach its full capacity. With the joining of co-occurrence ($\lambda>0$), impressive improvements are achieved, and SNSRec gradually reaches the peak (when $\lambda$ is around 0.4) of quality performance. These improvements are obvious evidence of co-occurrence pattern's importance for SNSRec (\textbf{C1}).  
As the involvement of co-occurrence becomes deeper ($\lambda>0.4$), the prediction performance declines. This is mainly because that an item's co-occurrence score is determined by calculating its co-occurrence patterns with all other items in $\mathbb{V}$. As the parameter value increases, the co-occurrence score becomes more susceptible to the influence of unrelated items, which in turn, adversely affects the final prediction scores. This adverse effect can potentially overshadow the relevance aspect of the predictions, resulting in a decrease in overall effectiveness. However, it is important to note that we cannot dismiss the role of co-occurrence due to the subsequent decline in performance, as the significant improvement in the earlier stage has already demonstrated the effectiveness and importance of co-occurrence pattern; 
(ii) \textbf{Importance of Preference}. We can treat the performances in the cases of $\lambda=0$ and $\lambda=1$ as the capabilities of preference and co-occurrence respectively, as only one of them is involved in SNSRec in the corresponding case. 
We can see that the preference representations dominate the quality performance of SNSRec, which is a quite common phenomenon in recommendation tasks \cite{kang2018self, huang2018improving}. 
This is also the reason why we mention that the co-occurrence pattern is used to complement items for prediction instead of playing a leading role. While co-occurrence information provides valuable insights, it is not as influential as preference representations in terms of improving the quality of predictions. This can further explain the reason of subsequent decline, when the $\lambda$ value becomes excessively large, the co-occurrence component gradually outweighs the preference component, leading to a decline in prediction performance. As the trends on CC with $\lambda$ is not quite obvious and similar trends can be found on other situations (different datasets and metrics), we omit corresponding presentations to save some space. 

\vspace{-2mm}
\subsubsection{Characteristic Analysis}
\begin{figure}
\centering
    \subfigure[targets number $B$]{
    \includegraphics[width=0.48\linewidth]{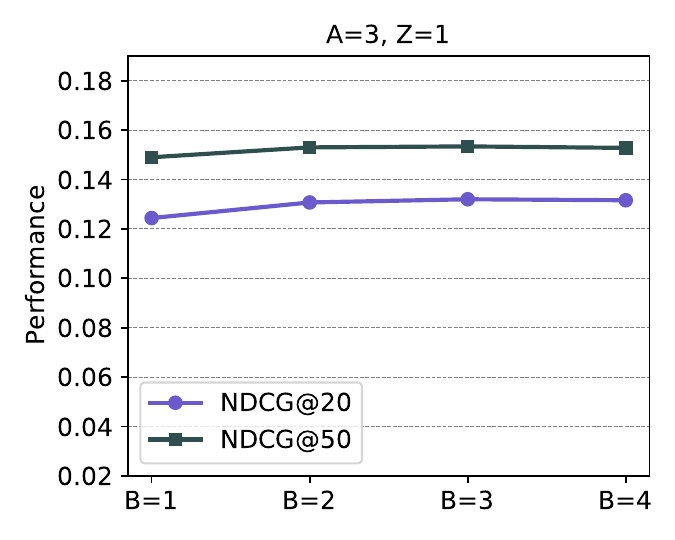}}
    \hfill
    \subfigure[previous sets length $A$]{
    \includegraphics[width=0.48\linewidth]{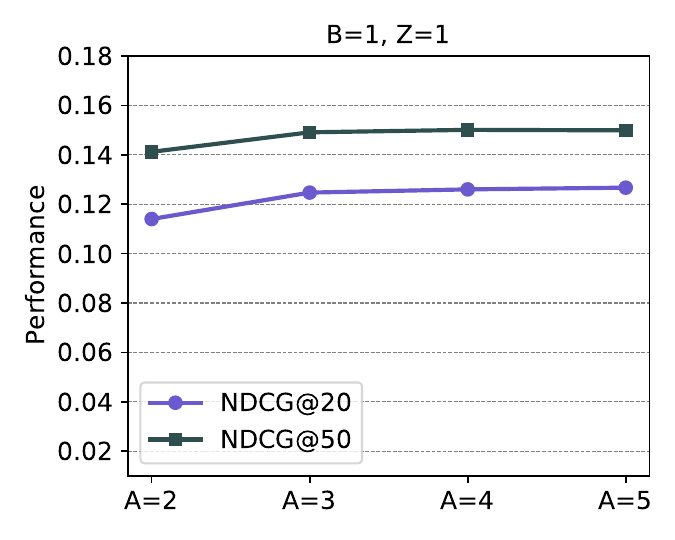}}
    \vspace{-3mm}
\caption{Performance trends \textit{w.r.t.} different sizes of $\mathcal{Y}^{(\boldsymbol{S})}$.}
    \label{figure-sdpp:dependency-trends}
    \vspace{-5mm}
\end{figure}

\Cref{figure-sdpp:dependency-trends} is used to analyze the characteristics of SNSRec, in which NDCG performance trends of Top-20 and Top-50 on TaoBao with varying $B$ and $A$ introduced in \Cref{euq-sdpp:SDPP-likelihood} are presented. A consistent trend can be observed in both sub-figures of \Cref{figure-sdpp:dependency-trends}, \textit{i.e.}, NDCG performances improve at first and then tend towards stability with the increase of $B$ and $A$ when fixing the other two factors ($A$ and $Z$ in \Cref{figure-sdpp:dependency-trends}(a), and $B$ and $Z$ in the other subfigure). 
This trend suggests that by considering more target sets or previous sets in our sets-level optimization criterion, more dependencies across the sets sequence can be captured, leading to substantial performance improvements. This effectively validates the importance of sequence dependencies for NSRec and demonstrates that our proposed optimization criterion is effective at capturing these dependencies.
To demonstrate the generality and applicability of SNSRec framework, results on different datasets displayed in this section are all obtained with $\lambda=0.2, B=1, A=3, \text{and } Z=1$.

\vspace{-1.7mm}
\section{RELATED WORK}
\vspace{-0.5mm}
\textbf{Temporal Sets Prediction.}
Unlike widely studied time series forecasting \cite{zhou2021informer, wu2021autoformer} and sequence prediction \cite{bengio2015scheduled, chiu2018state}, temporal sets prediction is proposed to untangle complicated relationships among items in the set and dependencies across temporal sets for predicting subsequent sets. 
Hu et al. \cite{hu2019sets2sets} first formulate this task and propose a set-to-set method for temporal sets prediction. To capture the dependencies across temporal sets, popular sequence-related models are employed, \textit{e.g.}, multi-heads self-attention \cite{sun2020dual} and RNN \cite{hu2019sets2sets}. Recently, some studies have attempted to propagate information between sets to model the complicated relationships. For example, Yu et al. 
\cite{yu2020predicting} perform graph convolutions on dynamic relationships graph to learn comprehensive set representations. Another study \cite{yu2022element} attempts to integrate collaborative signals into temporal sets by propagating element-guided message in a graph format. Although this work claims that the propagation is on set-level, the user-set interaction graph is actually constructed based on interactions between individual items and users instead of directly via set to set. In addition, previous TSP studies still use the item-level loss function (\textit{e.g.}, weighted mean square loss \cite{hu2019sets2sets} and binary cross-entropy loss \cite{yu2022element, sun2020dual} ) for training, in which the complicated relationships among sets are ignored. 

\vspace{-0.2mm}
\textbf{Next Basket Recommendation.}
Next Basket Recommendation is a relatively complex task compared to next item recommendation. This research problem originates from real-world shopping scenarios where users purchase a batch of items at a time. Initial studies primarily employed traditional methods such as markov chains \cite{rendle2010factorizing} and hierarchical representation model \cite{wang2015learning}. In recent years, sequence learning approaches based on deep learning have become mainstream, including the application and expansion of RNN-based \cite{yu2016dynamic}, attention-based \cite{bai2018attribute, zhang2020correlation}, neural graph-based \cite{liu2020basconv}, and hybrid \cite{zhang2020correlation, liu2022next} methods, which have further enhanced the effectiveness of NBRec. Additionally, the integration of contrastive learning in basket recommendation has recently garnered attention \cite{he2023robust, qin2021world}.

\vspace{-0.2mm}
\textbf{DPP for Diversification.}
Determinantal point process has been a trending research topic in machine learning area, largely because of the repulsive property and efficient algorithms. 
Applying the maximum a posteriori (MAP) to generate DPP samples as suggestions is the most common and direct manner \cite{han2017faster, gan2020enhancing}. However, MAP is NP-hard and is computationally expensive even using the popular greedy algorithm. Chen et al. \cite{chen2018fast} develop a novel algorithm to accelerate the MAP inference for DPP, which contributes to a flourishing phenomenon of employing DPP MAP in traditional Top-N recommendation models for generating diverse results, such as the studies \cite{wu2019pd, liu2020diversified}. 
Besides using MAP inference, normalized probability and approximate likelihood of DPP distribution are also studied for working as an optimization criterion for item recommendation \cite{liu2022determinantal, liu2024learning} and basket completion \cite{warlop2019tensorized}, respectively. 
There also exist multiple models considering using DPP for diversification of sequence data, \textit{e.g.}, addressing the bundle list recommendation problem by the sequence generation approach \cite{bai2019personalized}, considering video summarization as a supervised DPP subset selection problem \cite{zheng2021k},  and learning a diversity sampling function to generate a diverse yet likely set of future trajectories \cite{yuan2019diverse}. 

\vspace{-1.4mm}
\section{Acknowledgments}
\vspace{-0.9mm}
This work is supported by high performance computing center of Qinghai University. 

\vspace{-1.6mm}
\section{CONCLUSION}
\vspace{-0.9mm}
In this work, we focus on addressing the problem of next set recommendation via untangling naturally complicated relationships based on means of SDPP. 
Two types of dependencies (\textit{i.e.}, within a set DinS and across sequence sets DacSs) are formulated in the specialized sets-level optimization framework. 
To effectively achieve this, we try to explicitly formulate temporal sets as SDPP structures by introducing a novel co-occurrence representation and a pre-learned diversity kernel. 
The integrated and sets-level optimization criterion contribute to the construction of the general SNSRec framework and to impressive performance on both accuracy and diversity. 
It would be interesting to extend our framework to other research problems that are constrained in terms of performance due to the structural complexity, \textit{e.g.}, video summarization \cite{gong2014diverse} and bundle list recommendation \cite{bai2019personalized}. 
The CORs show impressive potential in our experiments, and it is worth the effort to explore further formulation and exploitation of COR for set-level prediction. 

\bibliographystyle{ACM-Reference-Format}
\bibliography{sample-base}

\end{document}